\documentclass[12pt,epsf]{article}
\usepackage{amsmath,amssymb,graphicx,epsfig,latexsym}
\numberwithin{equation}{section}
\newcommand{\be}{\begin{equation}}
\newcommand{\ee}{\end{equation}}

\newcommand{\beqa}{\begin{eqnarray}}
\newcommand{\eeqa}{\end{eqnarray}}
\newcommand{\nn}{\nonumber}

\def\boxit#1{\vbox{\hrule\hbox{\vrule\kern8pt
\vbox{\hbox{\kern8pt}\hbox{\vbox{#1}}\hbox{\kern8pt}}
\kern8pt\vrule}\hrule}}
\def\mathboxit#1{\vbox{\hrule\hbox{\vrule\kern8pt\vbox{\kern8pt
\hbox{$\displaystyle #1$}\kern8pt}\kern8pt\vrule}\hrule}}

\def\IB{\relax\hbox{$\inbar\kern-.3em{\rm B}$}}
\def\IC{\relax\hbox{$\inbar\kern-.3em{\rm C}$}}
\def\ID{\relax\hbox{$\inbar\kern-.3em{\rm D}$}}
\def\IE{\relax\hbox{$\inbar\kern-.3em{\rm E}$}}
\def\IF{\relax\hbox{$\inbar\kern-.3em{\rm F}$}}
\def\IG{\relax\hbox{$\inbar\kern-.3em{\rm G}$}}
\def\IGa{\relax\hbox{${\rm I}\kern-.18em\Gamma$}}
\def\IH{\relax{\rm I\kern-.18em H}}
\def\IK{\relax{\rm I\kern-.18em K}}
\def\IL{\relax{\rm I\kern-.18em L}}
\def\IP{\relax{\rm I\kern-.18em P}}
\def\IR{\relax{\rm I\kern-.18em R}}
\def\IZ{\relax\ifmmode\mathchoice
{\hbox{\cmss Z\kern-.4em Z}}{\hbox{\cmss Z\kern-.4em Z}}
{\lower.9pt\hbox{\cmsss Z\kern-.4em Z}} {\lower1.2pt\hbox{\cmsss
Z\kern-.4em Z}}\else{\cmss Z\kern-.4em Z}\fi}

\def\II{\relax{\rm I\kern-.18em I}}

\def\CA {{\cal A}}

\def\CD {{\cal D}}

\def\CP {{\cal P}}

\begin{document}

\setlength{\baselineskip}{7mm}
\begin{titlepage}

\begin{flushright}

{\tt NRCPS-HE-47-2012} \\

\end{flushright}

\vspace{1cm}
\begin{center}

{\Large \it  Non-Abelian  Tensor Gauge Fields \\
and\\
New Topological Invariants \\
\vspace{0,3cm}

} 

\vspace{1cm}
{ \it{George Georgiou} }
and
{ \it{George  Savvidy  } }\\
\vspace{1cm}
{\sl Demokritos National Research Center\\
Institute of Nuclear and Particle Physics\\
Ag. Paraskevi, GR-15310 Athens,Greece  \\
\centerline{\footnotesize\it E-mail: georgiou@inp.demokritos.gr, savvidy@inp.demokritos.gr}
}
\end{center}
\vspace{25pt}

\centerline{{\bf Abstract}}

\vspace{10pt}

\noindent

In this article we shall consider the tensor gauge fields which
are possible to embed into the existing framework of generalized YM theory
and therefore allows to construct the gauge invariant and metric independent
forms in  2n+4 and 2n+2 dimensions.
These new forms  are analogous to the   Pontryagin-Chern-Simons densities
in YM gauge theory and to the corresponding  series of densities in
2n+3 dimensions constructed recently in arXiv:1205.0027.

\begin{center}
\end{center}

\vspace{150 pt}


\end{titlepage}

\newpage

\pagestyle{plain}

\section{\it Introduction}

The Abelian and non-Abelian {\it chiral anomalies} can be determine
by a differential geometric method without having to evaluate the
Feynman diagrams \cite{Adler:1969gk,Bell:1969ts,Bardeen:1969md,Wess:1971yu,
Frampton:1983nr,Zumino:1983ew,Stora:1983ct,Faddeev:1984jp,Faddeev:1985iz,Mickelsson:1983xi,LBL-16443,
Manes:1985df,Treiman:1986ep,Faddeev:1987hg,DKFaddeev,AlvarezGaume:1985ex}.
The non-Abelian anomaly  in $2n-2$-dimensional space-time may be obtained  from
the Abelian anomaly in $2n$ dimensions by a series of reduction (transgression) steps
and can be represented in a compact integral form
\cite{Zumino:1983ew,Stora:1983ct,Faddeev:1984jp,Faddeev:1985iz,LBL-16443,Manes:1985df,
Treiman:1986ep,Faddeev:1987hg,AlvarezGaume:1985ex}.
In $\CD = 2n$ dimensions,  the $U_A(1)$ anomaly is given by a $2n$-form:
\be
\CP_{2n}=Tr(G^n)= d ~\omega_{2n-1},
\ee
where $\omega_{2n-1}$ is the  Chern-Simons form to $2n-1$ dimensions
\cite{Zumino:1983ew,Treiman:1986ep}:
\be\label{integralformforAbelian}
\omega_{2n-1}(A)= n \int^1_0 d t ~ Tr(AG^{n-1}_t),
\ee
where $G=dA +A^2$  is the 2-form  YM field-strength tensor
of the 1-form vector field $A = -ig A^{a}_{\mu} L_a dx^{\mu}$ and
$G_{t}= t G +(t^2-t)A^{2}$.

In the recent articles \cite{arXiv:1001.2808,Antoniadis:2012ep} the authors
have found the similar invariants in
non-Abelian tensor gauge field theory \cite{Savvidy:2005fi,Savvidy:2005zm,Savvidy:2005ki}. These forms
are defined in dimensions $\CD=2n+3 $~:
\be
\Gamma_{2n+3}(A,A_2)= Tr(G^n G_3) = d ~\sigma_{2n+2},
\ee
where $G_3=dA_2 +[A,A_2]$ is the 3-form  field-strength tensor
for the rank-2 gauge field
$A_2= -ig A^{a}_{\mu\lambda} L_a dx^{\mu} \wedge dx^{\lambda}$ and $G_{3t}=t G_3 +(t^2-t)[A,A_2]$.
The $(2n+2)$-form  $\sigma_{2n+2}$  is \cite{Antoniadis:2012ep}:
\be\label{sigma2n2}
\sigma_{2n+2}(A,A_2) =  \int^1_0 d t ~Tr(A  G^{n-1}_{t} G_{3t} +...+G^{n-1}_{t}  A G_{3t} + G^{n}_{t} ~ A_2).
\ee

The very fact that the tensor gauge fields introduced in
\cite{Savvidy:2005fi,Savvidy:2005zm,Savvidy:2005ki} are symmetric
over their last indices (see equation (\ref{symmertic}) ) prevents the
construction of the invariant forms involving higher rank tensor
gauge fields, that is the fields of the rank higher than two.
Our intension in this article is to demonstrate that a class of the
non-Abelian tensor gauge fields (\ref{symmertic})
can and should be extended to include new fields
and therefore allows to construct the invariant forms in $\CD=2n+4 $ and $\CD=2n+2$ dimensions.
These new forms $\Phi_{2n+4} =d \psi_{2n+3}$ and $\Omega_{2n+2} =  d \chi_{2n+1}$
are analogous to the   Pontryagin-Chern-Simons densities
$\CP_{2n}=d \omega_{2n-1}$  in
YM gauge theory and to the corresponding  series of densities
$\Gamma_{2n+3}=d \sigma_{2n+2}$ found in \cite{Antoniadis:2012ep}.

In the next section we shall introduce the {\it tensor gauge fields
which are possible to embed into the existing framework of generalized YM theory},
their gauge transformations and the corresponding
field strength tensor. In the third and fourth sections the invariant forms
$\Phi_{2n+4}$ and $\Omega_{2n+2}$ will be constructed. In the fifth section
we shall consider the most general tensor gauge fields from a geometrical
point of view \cite{Savvidy:2005fi,Savvidy:2005zm,Savvidy:2005ki}.

\section{\it Non-Abelian  Tensor Gauge Fields}

In the model of massless tensor gauge fields
suggested in \cite{Savvidy:2005fi,Savvidy:2005zm,Savvidy:2005ki}
the gauge fields are defined as rank-$(s+1)$ tensors
\be\label{symmertic}
A^{a}_{\mu\lambda_1 ... \lambda_{s}}(x),
\ee
which are totally symmetric with respect to the
indices $  \lambda_1 \dots\lambda_{s}  $. The number of symmetric
indices $s$ runs from zero to infinity. {\it A priori} the tensor fields
have no symmetries with respect to the first index  $\mu$.
The index $a$ corresponds to the generators $L_a$
of an appropriate Lie algebra.
The extended non-Abelian gauge transformation $\delta_{\xi} $
of the tensor gauge fields is defined as
\beqa\label{polygauge}
\delta_{\xi}  A_{\mu} &=& \partial_{\mu}\xi -i g[A_{\mu},\xi] \\
\delta_{\xi}  A_{\mu\lambda} &=& \partial_{\mu}\xi_{\lambda} -i g[A_{\mu},\xi_{\lambda}]
-i g [A_{\mu\lambda},\xi]\nonumber\\
\delta_{\xi}  A_{\mu\lambda_1\lambda_2} &=& \partial_{\mu}\xi_{\lambda_1\lambda_2}
-i g[A_{\mu},\xi_{\lambda_1\lambda_2}]-
i g[A_{\mu\lambda_1},\xi_{\lambda_2}]-i g [A_{\mu\lambda_2},\lambda_1 ]
-i g [A_{\mu\lambda_1\lambda_2},\xi],\nn\\[-7pt]
 &\cdot&\nn\\[-17pt]
 &\cdot&\nn\\[-17pt]
 &\cdot&\nn
\eeqa
where $\xi^{a}_{\lambda_1 ... \lambda_{s}}(x)$ are totally symmetric gauge parameters
and comprises a closed algebraic structure. The generalized field-strength tensors
are defined as
\cite{Savvidy:2005fi,Savvidy:2005zm,Savvidy:2005ki}:
\beqa\label{fieldstrengthparticular}
G_{\mu\nu} &=&
\partial_{\mu} A_{\nu} - \partial_{\nu} A_{\mu} -
i g [A_{\mu}~A_{\nu}],\\
G_{\mu\nu,\lambda} &=&
\partial_{\mu} A_{\nu\lambda} - \partial_{\nu} A_{\mu\lambda} -
i g  (~[A_{\mu}~A_{\nu\lambda}] + [A_{\mu\lambda}~A_{\nu}] ~),\nn\\
G_{\mu\nu,\lambda\rho} &=&
\partial_{\mu} A_{\nu\lambda\rho} - \partial_{\nu} A_{\mu\lambda\rho} -
i g (~[A_{\mu}~A_{\nu\lambda\rho}] +
 [A_{\mu\lambda}~A_{\nu\rho}]+[A_{\mu\rho}~A_{\nu\lambda}]
 + [A_{\mu\lambda\rho}~A_{\nu}] ~),\nn\\[-7pt]
 &\cdot&\nn\\[-17pt]
 &\cdot&\nn\\[-17pt]
 &\cdot&\nn
\eeqa
and transform homogeneously
with respect to the extended gauge transformations $\delta_{\xi} $.
The tensor gauge fields are in the matrix representation
$A^{ab}_{\mu\lambda_1 ... \lambda_{s}} =
(L_c)^{ab}  A^{c}_{\mu\lambda_1 ... \lambda_{s}} = i f^{acb}A^{c}_{\mu
\lambda_1 ... \lambda_{s}}$  with
$f^{abc}$ - the structure constants of the Lie algebra.

Using field-strength tensors one can construct  infinite series of forms
$
{{\cal L}}_{s}
$
invariant under the
transformations $\delta_{\xi} $. They are quadratic in field-strength
tensors. The
first terms are given by the formula
\cite{Savvidy:2005fi,Savvidy:2005zm,Savvidy:2005ki}:
\beqa\label{totalactiontwo}
{{\cal L}}=  {{\cal L}}_{YM}  +  {{\cal L}}_2 + ... =
&-&{1\over 4}G^{a}_{\mu\nu}G^{a}_{\mu\nu}\nn\\
&-&{1\over 4}G^{a}_{\mu\nu,\lambda}G^{a}_{\mu\nu,\lambda}
-{1\over 4}G^{a}_{\mu\nu}G^{a}_{\mu\nu,\lambda\lambda}\nn\\
&+&{1\over 4}G^{a}_{\mu\nu,\lambda}G^{a}_{\mu\lambda,\nu}
+{1\over 4}G^{a}_{\mu\nu,\nu}G^{a}_{\mu\lambda,\lambda}
+{1\over 2}G^{a}_{\mu\nu}G^{a}_{\mu\lambda,\nu\lambda}+...
\eeqa
The Lagrangian
contains quadratic in gauge fields kinetic terms, as well as cubic and
quartic terms  describing
non-linear interactions of gauge fields with dimensionless
coupling constant $g$.

Here we shall consider the set of fields which can be embedded into the existing framework
of generalized YM theory
and {\it should be unified with the previous system of fields} (\ref{symmertic}). First
let us consider the rank-3 field which is now antisymmetric over its last two indices
\be
A^{a}_{\mu \sigma_1 \sigma_{2}}(x) = - A^{a}_{\mu \sigma_{2} \sigma_1 }(x),
\ee
for this field the gauge transformation should be defined in following way
\beqa\label{antipolygauge}
\delta A_{\mu} &=&   \partial_{\mu}\xi
-ig [A_{\mu} , \xi] ,~~~~~\nn\\
\delta A_{\mu \sigma_1 \sigma_2}& =&   \partial_{\mu}
-ig [ A_{\mu}, \zeta_{\sigma_1 \sigma_2}] -
ig [A_{\mu \sigma_1 \sigma_2},\xi] ,
\eeqa
where the tensor gauge parameter is  antisymmetric
$\zeta^{a}_{\sigma_1 \sigma_2}=-\zeta^{a}_{\sigma_2 \sigma_1 }$.
As one can verify these transformations form a closed algebra because
$$
[\delta_{\zeta},\delta_{\varphi}]A_{\mu\nu\lambda} =\delta_{\chi}A_{\mu\nu\lambda},
$$
where
\beqa\label{commutatorofparameterslow}
\chi =[\zeta,\varphi], ~~~~
\chi_{\sigma_1 \sigma_2} = [\zeta,\varphi_{\sigma_1 \sigma_2}]+[\zeta_{\sigma_1 \sigma_2},\varphi].
\eeqa
It is useful to compare the gauge transformations of the $A^{a}_{\mu\lambda_1 \lambda_2}$
in (\ref{polygauge}) and of the $A^{a}_{\mu \sigma_1 \sigma_2}$ in (\ref{antipolygauge}).
As one can see they formally coincide, expect the term  $ [A_{\mu  \lambda_1 } \xi_{\lambda_2 }] +
[A_{\mu \lambda_2 } \xi_{\lambda_1 }]$, which is explicitly symmetric
under $\lambda_1  \leftrightarrow \lambda_2$ permutations. Therefore these fields can
and should be unified into a general rank-3 gauge field $A^{a}_{\mu\nu\lambda}$ of which
 the symmetric and antisymmetric parts,
with respect to the last two indices, reproduce the original fields.

In general we shall define the infinite set of fields
\be\label{antisymmetricfield}
A^a_{\mu ~\sigma_1 \sigma_2 ~\rho_1 \rho_2~ ....~ \kappa_{ 1} \kappa_{2}}(x)
\ee
which are antisymmetric with respect to the permutations of pairs of indices
$$
\sigma_1 \leftrightarrow\sigma_2 ,~~~  \rho_1 \leftrightarrow \rho_2 ,~~~...,~~~
 \kappa_{ 1} \leftrightarrow \kappa_{2},
 $$
but are symmetric under any permutations of  these pairs
$$
\sigma_1 \sigma_2 \leftrightarrow \rho_1 \rho_2 ~,~~~ ....,~ \sigma_1 \sigma_2 \leftrightarrow
\kappa_{ 1} \kappa_{2} ,...
$$
Thus for these fields take place the following relations
\beqa
&A^a_{\mu ~\sigma_1 \sigma_2 ~\rho_1 \rho_2~ ....~ \kappa_{ 1} \kappa_{2}}=
-A^a_{\mu ~\sigma_2 \sigma_1  ~\rho_1 \rho_2~ ....~ \kappa_{ 1} \kappa_{2}}=
-A^a_{\mu ~\sigma_1 \sigma_2 ~\rho_2\rho_1 ~ ....~ \kappa_{ 1} \kappa_{2}}=...\nn\\
&-A^a_{\mu ~\sigma_1 \sigma_2 ~\rho_1 \rho_2~ ....~ \kappa_{2} \kappa_{ 1} }=
A^a_{\mu ~\rho_1 \rho_2~ \sigma_1 \sigma_2 ~ ....~ \kappa_{ 1} \kappa_{2}}=...
=A^a_{\mu ~ \kappa_{ 1} \kappa_{2}~\rho_1 \rho_2~ ....~ \sigma_1 \sigma_2}=...\nn\\
&=A^a_{\mu ~\sigma_1 \sigma_2 ~\kappa_{ 1} \kappa_{2} ~ ....~ \rho_1 \rho_2}=... \nn
\eeqa
To simplify the description of these fields one can say that we have the gauge
fields of the same type as in (\ref{symmertic}), but now the indices $\{\lambda \}$
are replaced by the symbols  $\{ \hat{\sigma} \}$ which are now the multi-indices
$$
\lambda \rightarrow \hat{\sigma}  \equiv  (\sigma^{'},  \sigma^{''}) .
$$
The new class of fields (\ref{antisymmetricfield}) can be represented in the form
\be\label{mixsymmetry}
A^a_{\mu \hat{\sigma}_1 ...\hat{\sigma}_s}(x),
\ee
these fields are totally symmetric under permutations of the symbols $\hat{\sigma}_i$
and are antisymmetric under permutation of the indices within the each symbol
 $\hat{\sigma}_i$. We shall define the gauge transformations of these fields as
\beqa\label{antipolygaugetransf}
\delta A^{a}_{\mu} &=&  \partial_{\mu} \xi
-ig [A_{\mu}, \xi] ,~~~~~\\
\delta A_{\mu\hat{\sigma}_1} &=&   \partial_{\mu}\zeta_{\hat{\sigma}_1}
-i g [A_{\mu},\zeta_{\hat{\sigma}_1}] -ig [A_{\mu\hat{\sigma}_1} ,\xi],\nonumber\\
\delta A_{\mu\hat{\sigma}_1 \hat{\sigma}_2}& =&  \partial_{\mu}\zeta_{\hat{\sigma}_1 \hat{\sigma}_2}
-ig [A_{\mu},\zeta_{\hat{\sigma}_1 \hat{\sigma}_2}] -
ig [ A_{\mu  \hat{\sigma}_1 }, \zeta_{\hat{\sigma}_2 }] -
ig [A_{\mu \hat{\sigma}_2 } ,\zeta_{\hat{\sigma}_1 }] -ig
[A_{\mu\hat{\sigma}_1 \hat{\sigma}_2},\xi],\nn\\
.........&.&............................\nn
\eeqa
where the gauge parameters
$$
\zeta^a_{\hat{\sigma}_1 ...\hat{\sigma}_s}
$$ are
totally symmetric under permutations of the symbols $\hat{\sigma}_i$
and are antisymmetric under permutation of the indices within the each symbol
 $\hat{\sigma}_i$. The field strength tensors are defined as follow
\beqa\label{fieldstrengthparticular}
G_{\mu\nu} &=&
\partial_{\mu} A_{\nu} - \partial_{\nu} A_{\mu} -
ig [~A_{\mu},~A_{\nu}],\\
G_{\mu\nu,\hat{\sigma}_1} &=&
\partial_{\mu} A_{\nu\hat{\sigma}_1} - \partial_{\nu} A_{\mu\hat{\sigma}_1} -
i g  (~[A_{\mu},~A_{\nu\hat{\sigma}_1}] + [A_{\mu\hat{\sigma}_1},~A_{\nu}] ~),\nn\\
G_{\mu\nu,\hat{\sigma}_1 \hat{\sigma}_2} &=&
\partial_{\mu} A_{\nu\hat{\sigma}_1 \hat{\sigma}_2}
- \partial_{\nu} A_{\mu \hat{\sigma}_1 \hat{\sigma}_2 } -
ig ( [A_{\mu}, A_{\nu\hat{\sigma}_1 \hat{\sigma}_2}] +
 [A_{\mu\hat{\sigma}_1}, A_{\nu\hat{\sigma}_2}]\nn\\
 &+&[A_{\mu\hat{\sigma}_2}, A_{\nu\hat{\sigma}_1}]
 + [A_{\mu\hat{\sigma}_1 \hat{\sigma}_2}, A_{\nu}]  ),\nn\\
 ......&.&............................................\nn
\eeqa
and are transforming   homogeneously with respect to the gauge transformations
$\delta_{\zeta} $ (\ref{antipolygaugetransf}).
In these multi-indices notation the above transformations are identical
with the original transformations (\ref{polygauge}) therefore one can define the invariant
Lagrangian, which formally
coincides with the one defined in \cite{Savvidy:2005fi,Savvidy:2005zm,Savvidy:2005ki},
by interchanging the $\{\lambda\}$'s by  $\{  \sigma \}$'s
\beqa\label{actionthree}
{{\cal L}^{'}}=&-&
  {1\over 4} G^{a}_{\mu\nu,\lambda \sigma}G^{a}_{\mu\nu,\lambda\sigma}
-{1\over 4}G^{a}_{\mu\nu}G^{a}_{\mu\nu,\lambda\sigma\lambda\sigma}\\
&+&{1\over 4}G^{a}_{\mu\nu,\lambda\sigma}G^{a}_{\mu\lambda,\nu\sigma}
+{1\over 4}G^{a}_{\mu\nu,\nu\sigma}G^{a}_{\mu\lambda,\lambda\sigma}
+{1\over 2}G^{a}_{\mu\nu}G^{a}_{\mu\lambda,\nu\sigma\lambda\sigma}+....\nn
\eeqa
The total Lagrangian is a sum of the (\ref{totalactiontwo}) and (\ref{actionthree}).
Our next intension is to show that in addition to the above gauge invariant
Lagrangian one can construct also the new metric independent densities which can be
added to the Lagrangian,  as it was in the lower dimensional
case \cite{arXiv:1001.2808}, and are relevant for the description of the
potential anomalies\footnote{
The tensor gauge fields considered in the proposed extension
of the Yang-Mills theory  (\ref{symmertic}) and (\ref{mixsymmetry}) are neither totaly
symmetric nor they are totaly antisymmetric,  they are of the
mixed-symmetry type.
The investigation of the tensor  fields of mixed-symmetry type were
made in relation with the string field theory, where the expansion
of the string field on the oscillators naturally introduces tensor fields of
mixed-symmetry \cite{Curtright:1979uz,Curtright:1980yj,Labastida:1986gy,
Labastida:1986gy,Aulakh:1986cb,Labastida:1987kw}. There is also extended
literature on the field-theoretical description
of mixed-symmetry free fields and cubic interaction
vertices between mixed-symmetry AdS fields within various
approaches \cite{Metsaev:1995re,Brink:2000ag,Metsaev:2002vr,Zinoviev:2002ye,
Alkalaev:2006rw,Boulanger:2008up,Campoleoni:2012th,Alkalaev:2010af,Boulanger:2011qt}.}.

\section{\it  New Invariant Densities in $2 n+4$ Dimensions}

In order to introduce higher-dimensional metric-independent densities it is convenient to use
the language of forms \cite{Zumino:1983ew,Stora:1983ct,Faddeev:1987hg}.
At the same time we have to stress that this language can not be used
in general description of these fields dynamics, because in the invariant Lagrangian
considered in the previous section all components of the fields are present. While
in the metric-independent densities only antisymmetric parts of the fields
are participating.
The one- and two-form gauge potentials are defined as
$A=-ig A^{a}_{\mu} L_a dx^{\mu}$ and $A_2=-ig A^{a}_{\mu\nu} L_a dx^{\mu} \wedge dx^{\nu}$
with the corresponding field-strength tensors (\ref{fieldstrengthparticular})
\footnote{One should keep in mind that the fields and field-strength tensors,
like $A^{a}_{\mu\sigma_1\sigma_2}$ and $G_{\mu\nu, \sigma_1\sigma_2}$, can not
be expressed in terms of forms, but only their
antisymmetric parts.}:
\be
G  = dA  + A^{2}  ~,~~~~G_3 = d A_2 +[A  , A_2].
\ee
The Bianchi identities 
are  of the form
\be
DG  =0,~~~~DG_3 +[A_2, G ]=0,
\ee
where $DG  = dG  + [A , G ]$ and $DG_3 = dG_3 + \{A, G_3\}$.
With the new gauge fields in hands we shall introduce the three-form gauge potential  as
$A_3=-ig A^{a}_{\mu\sigma_1\sigma_2} L_a dx^{\mu} \wedge dx^{\sigma_1}\wedge dx^{\sigma_2}$
and the corresponding field strength tensor (\ref{fieldstrengthparticular})
\be
 G_4 = d A_3 +\{ A  , A_3 \}
\ee
with the Bianchi identity
\be
DG_4 + [A_3, G]=0,
\ee
where $DG_4 = dG_4 +[A,G_4]$.

Let us consider a higher-dimensional invariant density in $2n+4$ space-time dimensions:
\be
\Phi_{2n+4} = Tr(G^n  G_4),
\ee
which is a natural generalization of the Chern-Pontryagin form  $\CP_{2n}=Tr(G^n)$.
By direct computation of the derivative one can prove that $\Phi_{2n+4}$ is an exact form:
\beqa
d \Phi_{2n+4} &=& Tr(dG  G^{n-1}  G_4 +...+G^{n-1}  dG  G_4 + G^{n} dG_4)\nn\\
&=&Tr((dG  +[A ,G ]) G^{n-1}  G_4 +...  +G^{n-1} (dG  +[A, G])G_4+
G^{n}  dG_4 -\nn\\
&-&[A ,G ]  G^{n-1}  G_4-...-G^{n-1} [A, G] G_4)=\nn\\
&=& Tr(DG  G^{n-1}  G_4 +...+G^{n-1}  DG  G_4 + G^{n} (dG_4 +[A,G_4])\nn\\
&=& Tr(G^{n} (DG_4 +[A_3,G ]))=0. \nn
\eeqa
In this calculation one must change sign when transmitting the differential d through an odd form
or commuting odd forms using the cyclic property of the trace, and use Bianchi identities as well.
According to Poincar\'e's lemma, this equation implies that $\Phi_{2n+4}$ can be
locally written as an exterior differential of a certain (2n +3)-form.
In order to find the form of which $\Phi_{2n+4}$ is the derivative we have to find
its variation, induced by the variation of the fields $\delta A$ and
$\delta A_3$:
$$
\delta G = D (\delta A) ,~~~~\delta G_4 = D(\delta A_3) + \{A_3,\delta A \}
$$
yielding a variation of $\Phi_{2n+4}$ which is a total derivative:
\beqa\label{variation1}
\delta \Phi_{2n+4} &=&\delta Tr( G^n  G_4)=
Tr(D\delta A G^{n-1}  G_4 +...+G^{n-1}  D\delta A G_4 + G^{n} D\delta A_3 +
G^n \{ A_3 \delta A\})\nn\\
&=&Tr(D\delta A G^{n-1}  G_4 +...+G^{n-1}  D\delta A G_4 + G^{n} D\delta A_3 +\nn\\
&+& \delta A ([ A_3, G] G^{n-1} + G[ A_3, G]G^{n-2}+...+
G^{n-1} [ A_3, G]))  =\nn\\
&=&Tr(D\delta A G^{n-1}  G_4 +...+G^{n-1}  D\delta A G_4 + G^{n} D\delta A_3 -\nn\\
&-& \delta A ~DG_4~ G^{n-1}  -...- \delta A ~G^{n-1}~ DG_4)  =\nn\\
&=& TrD ( \delta A  G^{n-1}  G_4 +...+G^{n-1}  \delta A G_4 + G^{n} \delta A_3  )=\nn\\
&=&d ~Tr(\delta A  G^{n-1}  G_4 +...+G^{n-1}  \delta A  G_4 + G^{n}  \delta A_3).
\eeqa
Following \cite{Zumino:1983ew}, we introduce
a one-parameter family of potentials and strengths  through the parameter t ($0\leq t \leq 1$):
\be
A_{t}= t A,~~~G_{t}= t G +(t^2-t)A^{2},~~~
A_{3t}= t A_3,~~~G_{4t}= t G_4 +(t^2-t)\{A,A_{3}\},
\ee
so that the equation (\ref{variation1}) can be rewritten as
$$
\delta Tr(G^n_t ~ G_{4t}) = d ~Tr(\delta A_t  G^{n-1}_t  G_{4t} +...
+G^{n-1}_t  \delta A_t  G_{4t}
+ G^{n}_t  \delta A_{3t}).
$$
With $\delta = \delta t (\partial/ \partial t)$ and $\delta A_{t}=A \delta t$,
$\delta A_{3t}=A_3 \delta t$
we shall get by integration the
desired result:
\be\label{result}
Tr(G^n  G_4) = d ~\psi_{2n+3}~,
\ee
where the corresponding secondary $(2n+3)$-form is
\be\label{secondaryformnew}
\psi_{2n+3}(A,A_3) =  \int^1_0 d t ~Tr(A  G^{n-1}_{t} G_{4t} +...
+G^{n-1}_{t}  A G_{4t} + G^{n}_{t} ~ A_3).
\ee
The expression (\ref{secondaryformnew}) can easily be evaluated for n=1 in five dimensions:
\be\label{sigma4}
\psi_{5} =\int^1_0 d t ~Tr(A  G_{4t}  + G_{t} ~ A_3)
= Tr( G A_3)
\ee
In seven dimensions, $n=2$, we have
$$
\psi_{7} =\int^1_0 d t ~Tr(A  G_{t} G_{4t} +G_{t}  A G_{4t} + G^{2}_{t} ~ A_3),
$$
and after integration over t we get a secondary 7-form:
\beqa\label{newsecondary}
\psi_{7}(A,G,G_4)&=&{1\over 3} Tr( A  G  G_4 + A  G_4 G   + A_3 G^{2}
- {1\over 2} A^3  G_4\nn\\
&-&{1\over 2} (A^2  A_3 +A A_3 A  + A_3 A^2 )G  + {1\over 2} A^4  A_3 ).
\eeqa
The new property of the last functional compared with $\psi_{5}$ is that when the
field-strength tensors tend to zero,
$G =G_4=0$,  the above form does not vanish and is equal to
\be\label{remnant}
 {1\over 6 } Tr(   A^4 A_3 ).
\ee
The subsequent forms $\psi_{2n+3}$ are in $\CD=2n+3=5,7,9,11,\dots$ dimensions.

\section{\it Invariant Forms in Eight and Ten Dimensions}

In this section we focus on a series of invariant forms that can be constructed in $2n+2$ dimensions.\\
We start with a special invariant form that can be constructed in eight dimensions
\be
\Omega_8= Tr(G  G_6 + G_4 G_4),
\ee
where the 6-form strength tensor (\ref{fieldstrengthparticular}) is
\be
G_6 = dA_5 + \{A, A_5\} + 2 A^2_3,~~~~~DG_6 +2[A_3, G_4] + [A_5,G]  =0
\ee
and their gauge transformations (\ref{fieldstrengthtensortransfor}) are
\be
\delta G_6 = [G_6,\xi] + 2[G_4,\zeta_2] +[G,\zeta_4], ~~~\delta G_4 = [G_4,\xi] +[G,\zeta_2].
\ee
Using the above formulas one can get convince that the 8-form $\Omega_8$ is gauge
invariant and is exact $d \Omega =0$. Thus we have
\be
\Omega_8 = d \chi_7, ~~~~~\chi_7 = Tr(G A_5 + G_4 A_3).
\ee
The next invariant form of a similar structure can be constructed in ten dimensions
\be
\Omega_{10}= Tr(G  G_8 + 3 G_4 G_6),
\ee
where the 8-form strength tensor is
\be
G_8 = dA_7 + \{A, A_7\} +3 \{A_3, A_5\},~~~~~DG_8 +3 [A_3, G_6] + 3 [A_5,G_4]+ [A_7,G]  =0
\ee
and the gauge transforms are
\be
\delta G_8 = [G_8,\xi] + 3[G_6,\zeta_2] +3[G_4,\zeta_4]+[G,\zeta_6].
\ee
It is straightforward to show that the form $\Omega_{10}$ is gauge invariant and is exact $d \Omega_{10} =0$. Thus
\be
\Omega_{10} =d \chi_9,~~~~\chi_9 = Tr(G A_7 +3 G_4 A_5 +{3\over 2}G_6 A_3)
\ee
The general form of these invariants can be written as
\be
\Omega_{2n+2} = Tr(G G_{2n} + \alpha_1 G_4 G_{2n-2} +\alpha_2 G_6 G_{2n-4}+....)= d \chi_{2n+1}
\ee
where
\be
\chi_{2n+1}= Tr(G A_{2n-1}+\beta_1 G_4 A_{2n-3}+\beta_2 G_{6} A_{2n-5}+...)~.
\ee
and $\alpha_i$, $\beta_i$ are certain numerical coefficients.
The forms $\chi_{2n+1}$ are defined in  $\CD=2n+1=5,7,9,11,\dots$ dimensions. It is also true that
$\Omega_{6}  \equiv \Phi_6$, see (\ref{result}) and (\ref{sigma4}).

\section{\it Most General Tensor Gauge Fields}

The tensor gauge fields (\ref{symmertic}) introduced in \cite{Savvidy:2005fi,Savvidy:2005zm,Savvidy:2005ki}
do have a geometrical interpretation if one introduce a unite tangent vector $e^{\mu}$ and
consider the gauge field $\CA_{\mu}(x,e)$  depending on this variable
and then define the extended gauge transformation as in \cite{Savvidy:2005ki}
\be\label{extendedgaugetransformation}
\CA^{'}_{\mu}(x,e) = U(\xi)  \CA_{\mu}(x,e) U^{-1}(\xi) -{i\over g}
\partial_{\mu}U(\xi) ~U^{-1}(\xi),
\ee
where the unitary transformation matrix
is given by the expression
$
U(\xi) = exp\{ig \xi(x,e) \}
$
and the gauge parameter $\xi(x,e)$ has the following expansion
\be
\xi(x,e)=   \sum^{\infty}_{s=0}~{1\over s!}
\xi^{a}_{\lambda_1 ... \lambda_{s}}(x) ~L_a   e^{\lambda_1}...e^{\lambda_s}.
\ee
Using this language one can consider the  fields  of mixed symmetry (\ref{mixsymmetry}) introduced
in this article as the gauge field depending also on antisymmetric wedge product
\be
\omega^{\hat{\sigma}} = e^{\sigma_1} \wedge e^{\sigma_2}
\ee
so that
\be
\CA_{\mu}(x,e,\omega  ).
\ee
It's expansion in $\omega$  will generate all fields considered in the previous chapters, but
in addition it will generate tenors fields which have both - vector $\{\lambda\}$ and
multi-indices $\{\hat{\sigma}\}$
\be\label{verygeneral}
A_{\mu, \lambda_1,\lambda_2,..., \hat{\sigma}_1,\hat{\sigma}_2,... }.
\ee
These fields are symmetric under any permutations of all
these indices and antisymmetric under permutations within each multi-index.
Because in four dimensions one can construct even higher rank independent wedge
products of the vector $e^{\mu}$, such as $e^{\sigma_1} \wedge e^{\sigma_2} \wedge e^{\sigma_3}$ and
$e^{\sigma_1} \wedge e^{\sigma_2} \wedge e^{\sigma_3} \wedge e^{\sigma_4}$, one can consider
tensor fields depending on these antisymmetric variables. If one use the multi-index variable
$ \hat{\sigma} $ to denote double $\hat{\sigma}  \equiv  (\sigma^{'},  \sigma^{''})  $, triple
$ (\sigma^{'},  \sigma^{''},  \sigma^{'''})$  and quadric
$ (\sigma^{'},  \sigma^{''},  \sigma^{'''},  \sigma^{''''})$ multi-indices then the tensor
gauge fields will be of the same form as in (\ref{verygeneral}), but with double, triple and quadric
multi-indices. It is not difficult to
find out their gauge transformations and corresponding field strength tensors.

\section{\it Conclusions}

In this article we demonstrated that a class of the
non-Abelian tensor gauge fields (\ref{symmertic}) considered in
\cite{Savvidy:2005fi,Savvidy:2005zm,Savvidy:2005ki}
{\it can and should be extended} to include new fields (\ref{antisymmetricfield})
and therefore allows to construct a metric independent and  gauge invariant forms
in $\CD=2n+4 $ and $\CD=2n+2$ dimensions.
These new forms $\Phi_{2n+4} =d \psi_{2n+3}$ and $\Omega_{2n+2} =  d \chi_{2n+1}$
are analogous to the   Pontryagin-Chern-Simons densities
$\CP_{2n}=d \omega_{2n-1}$  in
YM gauge theory and to the corresponding  series of densities
$\Gamma_{2n+3}=d \sigma_{2n+2}$ found in \cite{Antoniadis:2012ep},
yielding the potential anomalies in gauge field theory. The above general considerations
should be supplemented  by an explicit calculation of loop diagrams involving chiral fermions.
The argument in favor of the existence of
these potential anomalies is based on the fact that they fulfill Wess-Zumino consistency conditions
\cite{Wess:1971yu,Zumino:1983ew,Stora:1983ct,Faddeev:1984jp,LBL-16443,Manes:1985df,Treiman:1986ep,
Faddeev:1987hg,AlvarezGaume:1985ex,Grimm:1974pr,Bardeen:1984pm,
AlvarezGaume:1984dr,AlvarezGaume:1985yb}.
At the same time, these invariant densities constructed on the space-time manifold
have their own independent value since they suggest the existence of  new
invariants characterizing topological properties of a manifold and can be added
to the invariant Lagrangian.

This work was supported in part by the General Secretariat for Research and
Technology of Greece and from the European Regional Development Fund (NSRF
2007-13 ACTION, KRIPIS).

\appendix
\section{ \it Field Strengths Transformations}

The gauge transformations $
\delta_{\xi} $ of non-Abelian tensor gauge fields
were defined in \cite{Savvidy:2005fi,Savvidy:2005zm,Savvidy:2005ki} as (\ref{polygauge}).
The generalized field-strength tensors
(\ref{fieldstrengthparticular}) transform homogeneously
under these gauge transformations $\delta_{\xi} $:
\beqa\label{fieldstrengthtensortransfor}
\delta G^{a}_{\mu\nu}&=&  -i g  [G_{\mu\nu} ~\xi] , \\
\delta G^{a}_{\mu\nu,\lambda} &=& -i g  (~[~G_{\mu\nu,\lambda}~ \xi ]
+  [G_{\mu\nu} ~\xi_{\lambda}]~),\nonumber\\
\delta G^{a}_{\mu\nu,\lambda\rho} &=& - i g
(~[G^{b}_{\mu\nu,\lambda\rho} ~\xi]
+ [ G_{\mu\nu,\lambda} ~\xi_{\rho}] +
[G_{\mu\nu,\rho} ~\xi_{\lambda}] +
[G_{\mu\nu} ~\xi_{\lambda\rho}]~),\nn\\[-7pt]
......&.&............................................\nn
\eeqa

\vfill

\end{document}